# ORBITAL ROTATION WITHOUT ORBITAL ANGULAR MOMENTUM: MECHANICAL ACTION OF THE SPIN PART OF THE INTERNAL ENERGY FLOW.

O. V. Angelsky[1], A. Ya. Bekshaev[2], P. P. Maksimyak[1],
A. P. Maksimyak[1], S. G. Hanson[3], C. Yu. Zenkova[4]

[1] Correlation Optics Department, Chernivtsi National University, 2, Kotsyubinsky Str., Chernivtsi 58012, Ukraine
[2] Physical Department, Odessa I.I. Mechnikov National University, Dvorianska 2, Odessa 65082, Ukraine
[3] DTU Fotonik, Department of Photonics Engineering, DK-4000 Roskilde, Denmark
[4] Department of Optics and Spectroscopy, Chernivtsi National University, 2, Kotsyubinsky Str., Chernivtsi 58012, Ukraine

*angelsky@itf.cv.ua*

**Abstract**: It is known that internal energy flow in a light beam can be divided into the orbital flow, associated with the macroscopic energy redistribution within the beam, and the spin flow originating from instantaneous rotation of the field vectors inherent in circular or elliptic polarization. In contrast to the orbital one, experimental observation of the spin flow constituent seemed problematic because (i) it does not manifest itself in the visible transformation of the beam profile and (ii) it converts into the orbital flow upon tight focusing of the beam, usually employed for the energy flow detection by the mechanical action on probe particles. We propose a two-beam interference technique that permits to obtain appreciable level of the spin flow in moderately focused beams and to detect the orbital motion of probe particles within a field where the transverse energy circulation is associated exclusively with the spin flow. This result can be treated as the first demonstration of mechanical action of the spin flow of a light field.

**References and links**

1. M. J. Padgett and L. Allen, "The Poynting vector in Laguerre–Gaussian laser modes," Opt. Commun. **121**, 36–40 (1995)
2. L. Allen and M. J. Padgett, "The Poynting vector in Laguerre–Gaussian beams and the interpretation of their angular momentum density," Opt. Commun. **184**, 67–71 (2000).
3. M. V. Vasnetsov, V. N. Gorshkov, I. G. Marienko and M. S. Soskin, "Wavefront motion in the vicinity of a phase dislocation: "optical vortex"," Opt. Spectrosc. **88**, 260–265 (2000).
4. V. A. Pas'ko, M. S. Soskin and M. V. Vasnetsov, "Transversal optical vortex," Opt. Commun. **198**, 49–56 (2001).
5. J. Lekner, "Phase and transport velocities in particle and electromagnetic beams," J. Opt. A: Pure Appl. Opt. **4**, 491–499 (2002).

6. J. Lekner, "Polarization of tightly focused laser beams," J. Opt. A: Pure Appl. Opt. **5**, 6–14 (2003).
7. H. F. Schouten, T. D. Visser and D. Lenstra, "Optical vortices near sub-wavelength structures," J. Opt. B: Quantum Semiclass. Opt. **6**, S404 (2004).
8. A. Ya. Bekshaev and M. S. Soskin, "Rotational transformations and transverse energy flow in paraxial light beams: linear azimuthons," Opt. Lett. **31**, 2199–2201 (2006).
9. I. Mokhun, A. Mokhun and Ju. Viktorovskaya, "Singularities of the Poynting vector and the structure of optical field," Proc. SPIE **6254**, 625409 (2006).
10. I. I. Mokhun, "Introduction to linear singular optics," in *Optical correlation techniques and applications* (Bellingham: SPIE Press PM168, 2007, pp. 1–132).
11. M. V. Berry, "Optical currents," J. Opt. A: Pure Appl. Opt. **11**, 094001 (2009).
12. A. Bekshaev, K. Bliokh and M. Soskin, "Internal flows and energy circulation in light beams," J. Opt. **13**, 053001 (2011).
13. R. Khrobatin, I. Mokhun and J. Viktorovskaya, "Potentiality of experimental analysis for characteristics of the Poynting vector components." Ukr. J. Phys. Opt. **9**, 182–186 (2008).
14. A. Ya. Bekshaev and M. S. Soskin "Transverse energy flows in vectorial fields of paraxial beams with singularities," Opt. Commun. **271**, 332–348 (2007).
15. A. Bekshaev and M. Soskin, "Transverse energy flows in vectorial fields of paraxial light beams," Proc. SPIE **6729**, 67290G (2007).
16. A. Ya. Bekshaev, "Spin angular momentum of inhomogeneous and transversely limited light beams," Proc. SPIE **6254**, 625407 (2006).
17. A. Bekshaev and M. Vasnetsov, "Vortex flow of light: "Spin" and "orbital" flows in a circularly polarized paraxial beam," in *Twisted Photons. Applications of Light with Orbital Angular Momentum* (Weinheim: Wiley-VCH, 2011, pp. 13–24).
18. A. Ya. Bekshaev, "Oblique section of a paraxial light beam: criteria for azimuthal energy flow and orbital angular momentum," J. Opt. A: Pure Appl. Opt. **11**, 094003 (2009).
19. A. Ya. Bekshaev, "Role of azimuthal energy flows in the geometric spin Hall effect of light," arXiv:1106.0982v1 [physics.optics] (6 Jun 2011).
20. A. Ya. Bekshaev, "Transverse energy flow and the "running" behaviour of the instantaneous field distribution of a light beam," arXiv:1108.0784 [physics.optics] (3 Aug 2011).
21. A. T. O'Neil, I. MacVicar, L. Allen and M. J. Padgett, "Intrinsic and extrinsic nature of the orbital angular momentum of a light beam," Phys. Rev. Lett. **88**, 053601 (2002).
22. V. Garces-Chavez, D. McGloin, M. D. Summers, A. Fernandez-Nieves, G. C. Spalding, G. Cristobal and K. Dholakia, "The reconstruction of optical angular momentum after distortion in amplitude, phase and polarization," J. Opt. A: Pure Appl. Opt. **6**, S235–S238 (2004).
23. Y. Zhao, J. S. Edgar, G. D. M. Jeffries, D. McGloin and D. T. Chiu, "Spin-to-orbital angular momentum conversion in a strongly focused optical beam," Phys. Rev. Lett. **99**, 073901 (2007).
24. M. Dienerowitz, M. Mazilu and K. Dholakia, "Optical manipulation of nanoparticles: a review," Journal of Nanophotonics **2**, 021875 (2008).
25. A. Ya. Bekshaev, O. V. Angelsky, S. V. Sviridova, and C. Yu. Zenkova, "Mechanical action of inhomogeneously polarized optical fields and detection of the internal energy flows," Advances in Optical Technologies **2011**, 723901 (2011).
26. T. A. Nieminen, A. B. Stilgoe, N. R. Heckenberg and H. Rubinsztein-Dunlop, "Angular momentum of a strongly focused Gaussian beam," J. Opt. A: Pure Appl. Opt. **10**, 115005 (2008).
27. O. V. Angelsky, N. N. Dominikov, P. P. Maksimyak, T. Tudor, "Experimental revealing of polarization waves", Appl. Opt. **38**, No 14, 3112–3117 (1999).
28. O. V. Angelsky, S. B. Yermolenko, C. Yu. Zenkova and A. O. Angelskaya, "Polarization manifestations of correlation (intrinsic coherence) of optical fields," Appl. Opt. **47**, No 32, 5492-5499 (2008).

29. O. V. Angelsky, M. P. Gorsky, P. P. Maksimyak, A. P. Maksimyak, S. G. Hanson, C. Yu. Zenkova, "Investigation of optical currents in coherent and partially coherent vector fields," Opt. Express **19**, 660–672 (2011).
30. A. Gerrard and J.M. Burch, *Introduction to matrix methods in optics* (London, Wiley-Interscience, 1975).
31. A. Ashkin, *Optical Trapping and Manipulation of Neutral Particles Using Lasers* (Singapore: Hackensack, NJ : World Scientific, 2006).
32. Xi-Lin Wang, Jing Chen, Yongnan Li, Jianping Ding, Cheng-Shan Guo, and Hui-Tian Wang, "Optical orbital angular momentum from the curl of polarization," Phys. Rev. Lett. **105**, 253602 (2010).
33. Shaohui Yan, Baoli Yao, and Ming Lei, "Comment on "Optical Orbital Angular Momentum from the Curl of Polarization"," Phys. Rev. Lett. **106**, 189301 (2011).

## 1. Introduction

Study of the internal energy flows is a rapidly developing branch of physical optics (see, e.g., Refs. [1–18]). The internal flows (optical currents) not only constitute an "energy skeleton" of a light field, which reflects important physical characteristics of its spatial structure. They have proven to be valuable instruments for investigation of fundamental dynamical and geometrical aspects of the light fields' evolution and transformations [1–12], provide a natural language for explaining the special features of singular fields [1–4, 7–15], fields with angular momentum [8,14–18] and for interpreting the effects of spin-orbit interaction of light [12,18,19]. As physically meaningful and universal parameters of light fields, they permit to disclose physical mechanisms of the beam transformation upon free and restricted propagation and offer attractive possibilities for characterization of arbitrary light fields [12].

In the usual case of a monochromatic electromagnetic field, the electric and magnetic vectors can be taken in forms $\text{Re}[\mathbf{E}\exp(-i\omega t)]$, $\text{Re}[\mathbf{H}\exp(-i\omega t)]$ with complex amplitudes $\mathbf{E}$ and $\mathbf{H}$ ($\omega$ is the radiation frequency). Then, the time-average energy flow density is expressed by the Poynting vector $\mathbf{S}$ or the electromagnetic momentum density $\mathbf{p}$ distributions

$$\mathbf{S} = c^2 \mathbf{p} = gc \, \text{Re}(\mathbf{E}^* \times \mathbf{H}) \tag{1}$$

($g = (8\pi)^{-1}$ in the Gaussian system of units, $c$ is the light velocity). The total field momentum density (1) can be subdivided into the spin (SMD) and orbital (OMD) parts, $\mathbf{p} = \mathbf{p}_S + \mathbf{p}_O$, according to which sort of the beam angular momentum they are able to generate [11,14,15]:

$$\mathbf{p}_S = \frac{g}{4\omega} \text{Im}\left[\nabla \times (\mathbf{E}^* \times \mathbf{E} + \mathbf{H}^* \times \mathbf{H})\right], \quad \mathbf{p}_O = \frac{g}{2\omega} \text{Im}\left[\mathbf{E}^* \cdot (\nabla)\mathbf{E} + \mathbf{H}^* \cdot (\nabla)\mathbf{H}\right] \tag{2}$$

where $\mathbf{E}^* \cdot (\nabla)\mathbf{E} = E_x^* \nabla E_x + E_y^* \nabla E_y + E_z^* \nabla E_z$. Peculiar properties of the SMD and OMD contributions (2) reflect specific features of the "intrinsic" rotation associated with the spin of photons ($\mathbf{p}_S$) and of the macroscopic energy transfer ($\mathbf{p}_O$) in a light field. The quantities (2) provide deeper insight into thin details of the light field evolution and allow one to describe interrelations between the spin and orbital degrees of freedom of light [12,15–17,20].

However, wide practical application of the internal flow parameters is hampered by difficulties in their experimental measurement and/or visualization. At present, only indirect procedures are available, e.g., via the Stokes polarimetry [13], where the energy flow pattern is calculated from the measured amplitude, phase and polarization data. In this context, possibilities coupled with the energy flow visualization via the motion of probe particles, suspended within an optical field, attract a special attention [21–25]. This technique relies on assumption that the force acting on a particle is proportional to the local value of the field momentum. Though with serious precautions [12,25], this assumption qualitatively justifies for the OMD $\mathbf{p}_O$ whereas even the physical explanation of how the spin momentum can be transferred from the field to a particle is not clear. For example, a circularly polarized beam, as well as any its fragment, carries the "pure" angular momentum that can cause spinning motion of the absorbing particle, but there is no clear understanding whether and how the translational or orbital motion can appear in this situation [12,17]. Besides, the spin flow does not manifest itself in the visible changes of the beam profile upon propagation [17]. Although recent calculations [25] suggest no significant differences in mechanical action of the SMD and OMD, a direct unambiguous verification of their mechanical equivalence (e.g., in ability to produce corresponding light pressure on material objects) is highly desirable [12,17].

In the present paper, we describe experimental observation of the polarization-dependent orbital motion of suspended probe particles in a transversely inhomogeneous beam with circular polarization where rotational action of the OMD is absent or negligible. To the best of our knowledge, these results can be considered as the first experimental evidence of the mechanical action of the spin momentum (spin energy flow) of a light beam.

## 2. Spin and orbital flows in paraxial beams

Let us consider a paraxial light beam propagating along axis $z$, and let the transverse plane be parameterized by coordinate axes $x$, $y$. The spatial distribution of the electric vector in this beam can be described as [12,14,17]

$$\mathbf{E} = \mathbf{E}_\perp + \mathbf{e}_z E_z = \exp(ikz)\left(\mathbf{u} + \frac{i}{k}\mathbf{e}_z \operatorname{div}\mathbf{u}\right) \tag{3}$$

where the slowly varying vector complex amplitude $\mathbf{u} = \mathbf{u}(x, y, z)$ is related to complex amplitudes of orthogonal polarization components of the field (3), $\mathbf{e}_z$ is the unit vector of longitudinal direction, $k = \omega c$ is the radiation wavenumber. In the circular-polarization basis

$$\mathbf{e}_\sigma = \frac{1}{\sqrt{2}}\left(\mathbf{e}_x + i\sigma \mathbf{e}_y\right)$$

($\mathbf{e}_x$, $\mathbf{e}_y$ are unit vectors of the transverse coordinates, $\sigma = \pm 1$ is the photon spin number, or helicity),

$$\mathbf{u} = \mathbf{e}_+ u_+ + \mathbf{e}_- u_-, \tag{4}$$

$u_\sigma \equiv u_\sigma(x, y, z)$ is the scalar complex amplitude of the corresponding circularly polarized component [17]. Then, by using 'partial' intensity and phase distributions,

$$I_\sigma(x, y, z) = cg|u_\sigma(x, y, z)|^2, \quad \varphi_\sigma = \frac{1}{2i}\ln\frac{u_\sigma}{u_\sigma^*}, \tag{5}$$

the SMD and the transverse part of the OMD (2) can be expressed as sums of contributions belonging to the orthogonal polarization components,

$$\mathbf{p}_S = \mathbf{p}_{+S} + \mathbf{p}_{-S}, \quad \mathbf{p}_{O\perp} = \mathbf{p}_{+O} + \mathbf{p}_{-O}, \tag{6}$$

where [12]

$$\mathbf{p}_{\sigma S} = -\frac{\sigma}{2\omega c}\left[\mathbf{e}_z \times \nabla_\perp I_\sigma\right] = \frac{\sigma}{2\omega c}\nabla_\perp \times (\mathbf{e}_z I_\sigma), \tag{7}$$

$$\mathbf{p}_{\sigma O} = \frac{g}{\omega}\operatorname{Im}\left(u_\sigma^* \nabla_\perp u_\sigma\right) = \frac{1}{\omega c}I_\sigma \nabla_\perp \varphi_\sigma \tag{8}$$

and $\nabla_\perp = \mathbf{e}_x(\partial/\partial x) + \mathbf{e}_y(\partial/\partial y)$ is the transverse gradient.

In particular, Eq. (7) means that in beams with homogeneous circular polarization but inhomogeneous intensity, the SMD circulates around the intensity extrema [12,14,17]. In contrast, the internal OMD (6) is directed along the transverse phase gradient, and it is not difficult to realize conditions where the OMD vanishes or distinctly differs from the spin contribution, e.g., by direction, so that both contributions can be easily separated in experiment.

## 3. Analysis of the experimental approach

Direct observation of the internal energy flows via the field-induced motion of probe particles within a collimated laser beam is generally difficult because the transverse light pressure associated with momentum densities (7) and (8) is rather weak at usual beam intensities. To enlarge the effect, in usual schemes [21–24] a cell with suspended particles is placed near the focal plane of the strongly focusing objective, in which the incident beam is efficiently squeezed. The high numerical aperture (NA) of the objective guarantees sufficient concentration of the light energy to provide noticeable mechanical action. However, high NA is unfavourable for the SMD investigation since tight focusing of a circularly polarized beam induces partial conversion of the initial spin flow into the orbital one [12,23] and, consequently, even if the mechanical action is observed, one cannot definitely exclude that it is caused by the conversion-generated OMD. To avoid this ambiguity, the focusing strength should not be high: in accordance with known data [26], the spin-orbital conversion is negligible (does not exceed 1%) at NA $\lesssim$ 0.2 (focusing angle $\theta \approx 11°$). Of course, this leads to certain loss in the energy concentration but it can be made without essential reduction of the focal-region SMD if lowering the intensity is compensated by increasing the beam inhomogeneity (see Eq. (7)).

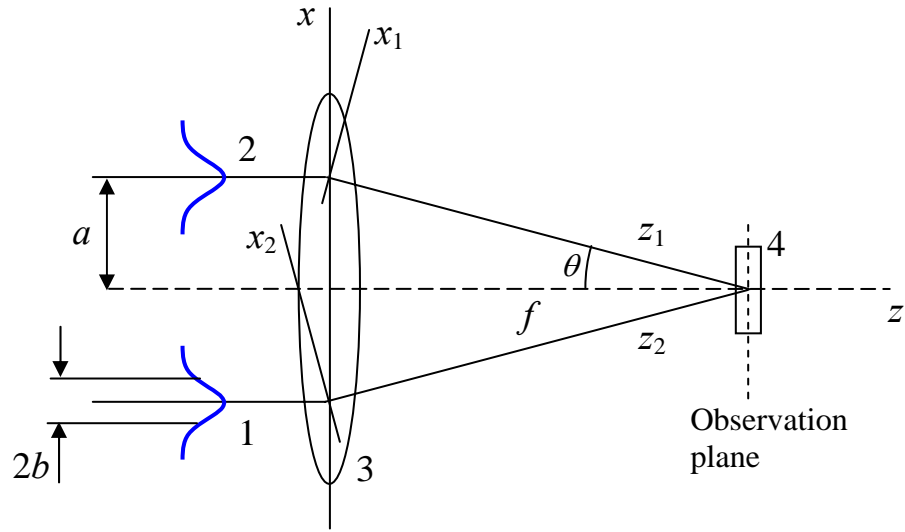

Fig. 1. Schematic of the experimental setup: (1), (2) input beams, (3) objective lens, (4) cell with probing particles suspended in the water. Axes $x_j$ and $z_j$ of the involved frames (see Eqs. (10)) are shown, axes $y_j$ are orthogonal to the figure plane.

To this end, the ideas of polarization interferometry developed in Refs. [27,28] can be employed (see Fig. 1). With this approach, the internal flows are studied in the field formed by superposition of two beams. Their polarizations (circular, elliptic, linear), phases, intensities and degree of mutual coherence can vary in wide ranges, which provides possibility to create diversity of optical fields with desirable properties [29]. By controlling the angle between the beams' axes, one may regulate the spatial intensity modulation (interference pattern) as well spatial inhomogeneity of polarization of the resulting field.

In our experiments, two identical beams obtained from a semiconductor laser ($\lambda = 0.67$ μm) with radii $b = 0.7$ mm (measured at the intensity level $e^{-1}$ of maximum) approach a microobjective with focal distance $f = 10$ mm. The beams are parallel to the objective axis and are located at $a = 1.3$ mm from it which provides the effective focusing angle $\theta = \arctan(a/f) \approx 7.4°$ and NA = 0.16; after focusing, they interfere in the focal region of the objective. If the beams are circularly polarized, they can be described by terms of Eqs. (4) – (8) corresponding to either $\sigma = +1$ or $-1$; once the helicity is fixed, the subscript $\sigma$ can be omitted from subsequent equations. Let both beams are Gaussian with the nominal input (just before the lens) complex amplitude distribution

$$u_j(x,y,z) = A_0 \exp\left(-\frac{\left[x+(-1)^j a\right]^2 + y^2}{2b^2}\right) \quad (j = 1, 2). \tag{9}$$

Behind the objective, each beam propagates along its own axis $z_j$ with focusing angle $\sim \arctan(b/f) \approx 0.07$ rad, which fairly corresponds to the paraxial regime. Therefore, in the proper coordinate frame $(x_j, y_j, z_j)$ (see Fig. 1), which is connected to the laboratory frame $(x, y, z)$ by relations

$$x_j = \left[x+(-1)^j a\right]\cos\theta - (-1)^j z\sin\theta, \quad y_j = y, \quad z_{1,2} = (-1)^j\left[x+(-1)^j a\right]\sin\theta + z\cos\theta, \tag{10}$$

its evolution is described by equation

$$u_j(x_j, y_j, z_j) = \eta A_0 \frac{\left(1-\frac{z_j}{f}\right) - i\frac{z_j}{z_R}}{\left(1-\frac{z_j}{f}\right)^2 + \left(\frac{z_j}{z_R}\right)^2} \exp\left\{-\frac{1}{2}(x_j^2 + y_j^2)\frac{\frac{1}{b^2} - \frac{ik}{z_R}\left[\frac{z_j}{z_R} - \frac{z_R}{f}\left(1-\frac{z_j}{f}\right)\right]}{\left(1-\frac{z_j}{f}\right)^2 + \left(\frac{z_j}{z_R}\right)^2}\right\} \tag{11}$$

where coefficient $\eta$ accounts for the energy losses in the focusing optical system, $z_R = kb^2$. Formula (11) can be readily derived from the common theory of Gaussian beams (see, e.g., Ref. [30]). Then, neglecting the small (in agreement to Eq. (3)) longitudinal components, the resulting amplitude distribution in the focal region can be found from equation

$$\left[u(x,y,z)\exp(ikz)\right]_{z=f+\delta} = u_1(x_1,y,z_1)\exp(ikz_1) + u_2(x_2,y,z_2)\exp(ikz_2) \tag{12}$$

where $\delta$ specifies the exact location of the observation plane with respect to the focus (in experiment, $\delta$ was adjusted empirically to provide the best conditions for particle trapping and manipulation), $z_j$ and $x_j$ should be replaced by their expressions (10) with allowance for $z = f + \delta$.

Properties of the interference pattern, calculated via Eqs. (7), (8) and (10) – (12) for conditions of Fig. 1, are illustrated by Fig. 2. It is seen that the circulatory flow of the spin nature exists within each lobe, while the OMD is, in fact, completely radial (this should be attributed to the beam divergence). This radial field momentum can be used for the probe particle confinement at a desirable off-center position [31], permitting to observe the SMD-induced orbital motion. Within an inhomogeneous optical field, any dielectric particle is subjected to the gradient force [24] that pulls the particle towards the intensity maximum (the beam axis); in contrast, the radial OMD of a divergent beam produces the radial light pressure that pushes the particle away from the axis. As a result, both forces can compensate each other at certain off-axial points within the central lobe of the interference pattern (e.g., points A and B in Fig. 2d), permitting stable trapping the particle at a position where azimuthal action of the SMD is the most efficient (compare Fig. 2d and Figs. 2a, b). In experiment, such conditions occur if the observation plane is located several microns behind the focus ($\delta > 0$).

Fig. 2c shows that due to strong intensity modulation, the SMD in the two-beam interference pattern is approximately 2.5 times higher than in a single Gaussian beam focused with the same NA objective. Noticeably, to reach the equivalent SMD level in a single Gaussian beam, conditions of NA $\approx$ 0.4 should be realized, when over 10% of the initial SMD would be transformed to the OMD [26]. The interference technique of the focal pattern formation permits to avoid this undesired conversion and to observe the mechanical action of the 'pure' spin flow without any contaminative influence of the orbital one.

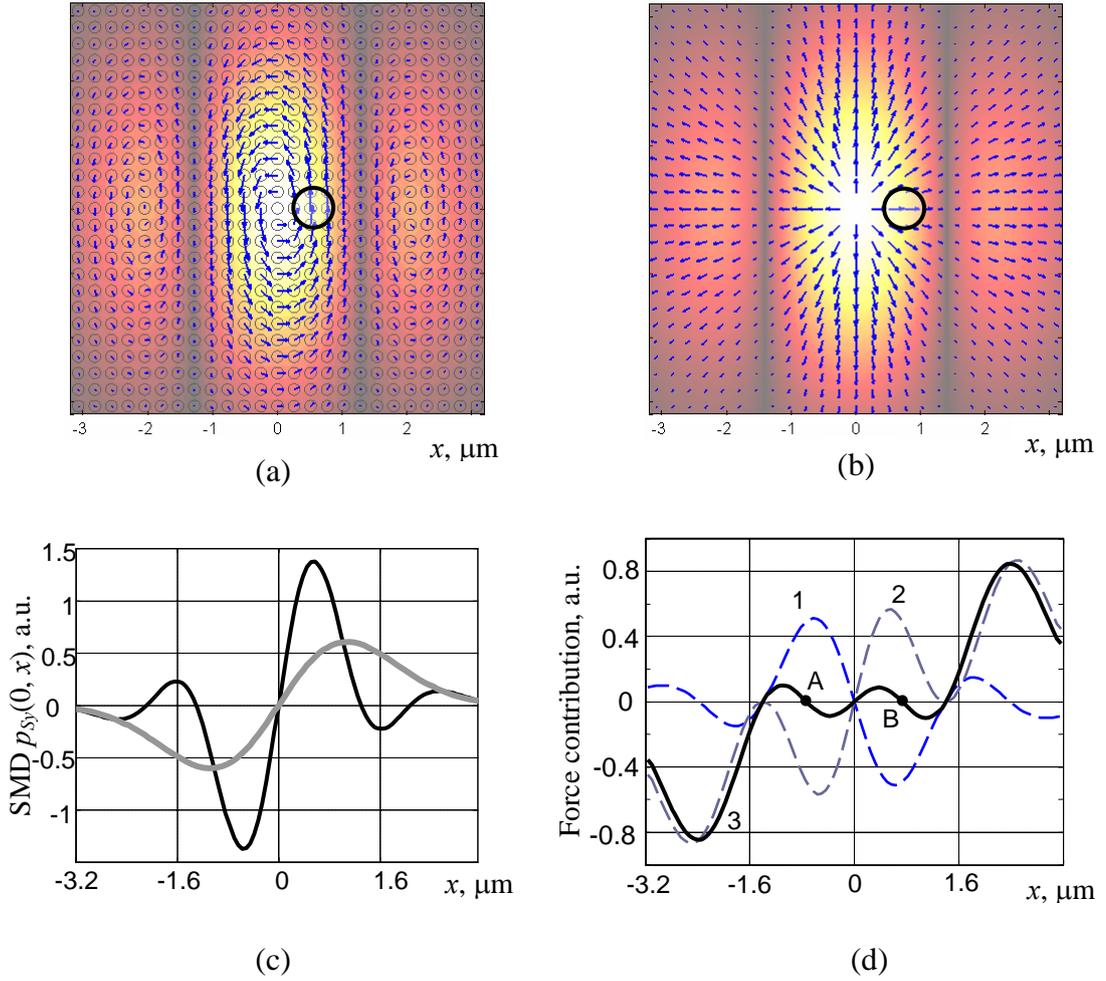

Fig. 2. Characteristics of optical field in the observation plane (see Fig. 1) for $\sigma = 1$, views against axis $z$. (a) SMD and (b) OMD maps (arrows) with the intensity distribution as a background; (c) actual SMD distribution along the $x$-axis (black curve) together with the SMD distribution for a single focused Gaussian beam with the same sum power (light curve); (d) qualitative pattern of the forces experienced by a probe particle at the $x$-axis: gradient force (curve 1), OMD-generated radial light pressure force (curve 2) and resulting force (black curve), A and B are points of stable equilibrium. In panel (a), polarization ellipses are shown on the background (because of small $\theta$, they have small eccentricities and visually look as circles); panels (a) and (b) also contain contours of a trapped particle (black circle) located at point B of panel (d).

## 4. Results and discussion

In experiment, the cell was used that contained ensemble of latex microparticles (refraction index 1.48) suspended in water. The particles were chosen so that their shape was close to ellipsoid with approximate size 1.5×1 μm, which was suitable for observing individual particles within a single lobe of the interference pattern formed in the focal region.

Experimental observations of the trapped particle motion in case when both superposed beams were circularly polarized, are represented by Fig. 3 and by the video clip in the Appendix. It is seen that the asymmetric particle spins around its own centre of mass, which is naturally explained by partial absorption of the incident circularly polarized light together with its inherent angular momentum. This effect is well known [21,24] and quite expectable in this situation. A new fact is that, simultaneously, the particle centre of mass evidently performs the orbital motion, which can only be associated with the azimuthal light pressure originating from the SMD circulation (see Fig. 2a). This attribution is confirmed by the rotation reversal when the input circular polarization changes the sign; besides, when both beams are linear, the particle stops.

Hence, the preliminary suggestion that the spin energy flow of an inhomogeneous circularly polarized beam can cause translational and orbital motion of probe particles, is experimentally verified. Among other things, this means that usual association "orbital motion of particles witnesses for the orbital angular momentum in the motive light field" is generally not correct, and possible contribution of the spin flow must be taken into account in experiments on the spin-to-orbital angular momentum conversion [23,26]. (To be true, in the known works treating this issue, in particular, Ref. [23], the spin flow action is absent or negligible, and their conclusions are quite truthful).

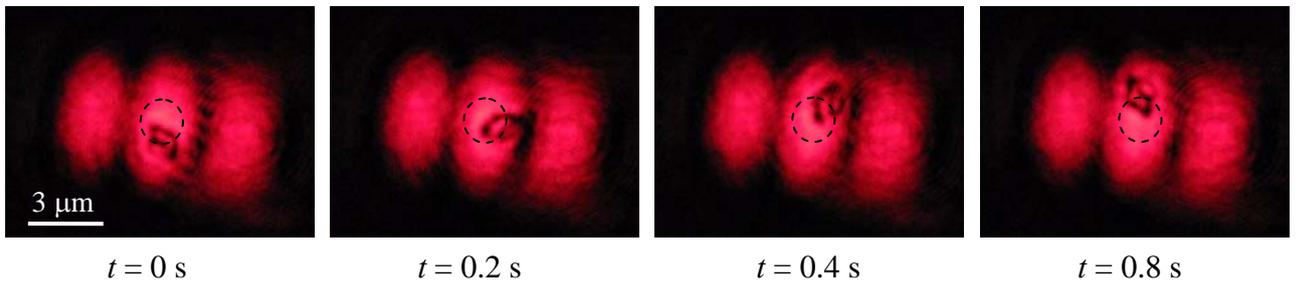

$t = 0$ s          $t = 0.2$ s          $t = 0.4$ s          $t = 0.8$ s

Fig. 3. Motion of a particle trapped within the central lobe of the interference pattern of Fig. 2. Frames represent consecutive positions of the particle (time labels below), dashed contour shows the particle trajectory. See also a video-clip in the Appendix.

It should be emphasized that beams with inhomogeneous intensity and uniform circular polarization, employed in this paper, are not unique examples of light field with nonzero SMD. In accordance with Eqs. (6), (7), quite similar SMD should appear in polarization-inhomogeneous beams. Such situations were recently discussed [32] but wrongly interpreted [33] as manifestations of a new category of the orbital angular momentum. Besides, high-NA focusing reported in Ref. [32] gives no certainty that the observed orbital motion of trapped particles is not caused by the OMD generated due to the spin-to-orbital conversion.

## 5. Conclusions

The results reported in this paper can be considered as the first, as far as we know, experimental evidence for the mechanical action of the spin momentum of light fields. This serves an additional confirmation for the mechanical equivalence of the spin and orbital momentum of light, despite the difference in their physical nature [12,16]. Additionally, we have demonstrated possibility of the SMD-induced particle transportation, which probably constitutes an interesting applied aspect of the observed phenomena. In our opinion, such a possibility opens up new promising opportunities for controllable optical manipulation procedures in which regulation and regime switching is realized via the polarization control alone, without change of the trapping beam intensity or spatial profile. Such techniques may be advantageous in many applications, e.g., when the high switching speed is important.

**Appendix.**

The orbital motion of a trapped particle in conditions corresponding to Fig. 3 is illustrated by the movie below.

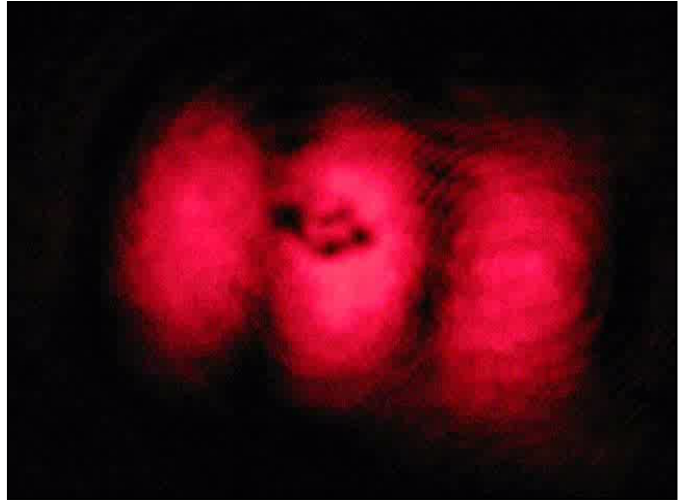